\documentclass[letterpaper,twocolumn,notitlepage]{article}
\usepackage{amsmath,amssymb,amsthm}
\usepackage{graphicx}
\usepackage{epstopdf}
\graphicspath{{images/}}

\begin{document}
\date{}
\title{\Large \bf High speed nanotribology with quartz crystal resonators\\
via Atomic Force Microscopy}
\author{{\rm $F. Z. Zhang^1$, $O. Marti^1$, $S. Walheim^2$, $T. Schimmel^{2,3}$}\\
$^1$Institute of Experimental Physics, Ulm University, 89069 Ulm, Germany\\
$^2$Institute of Nanotechnology, Forschungszentrum Karlsruhe, 76021 Karlsruhe\\
$^3$Institute of Applied Physics, University of Karlsruhe (TH),76131 Karlsruhe}
\renewcommand{\today}{Feb,23 2009}

\twocolumn[
  \begin{@twocolumnfalse}
  \maketitle

   \begin{abstract}

Friction measurements in the range of several meters per second are still of great interests. With the atomic force microscopy (AFM), the oscillation situation of the quartz crystal resonators of 3MHz resonance frequency are studied. And the oscillation speed could reach up to several m/s. Then the friction measurements are carried out on the Fischer Pattern, which is prepared on the quartz crystal resonator. In this article, we present how to measure the oscillation speeds of the quartz crystal resonator and the friction measurements under different designs. Calibration method and the calculation methods are also discussed in details. Although there are errors in the AFM measurements due to the setup itself and the oscillating quartz which could highly affect the sharpness of the cantilever tip, the results indicate that the local friction coefficients with oscillation are higher than that without oscillation.  \\
\\
 \end{abstract}
 \end{@twocolumnfalse}
]

\section{Introduction}
Friction is an everyday phenomenon in our life. Since a couple of hundreds of years, many researchers were trying to understand the mechanism of friction. Three laws of macroscopic friction had been discovered and well established\cite{Dowson} by Leonardo da Vinci\cite{Vinci}, Guillaume Amontons\cite{Amontons} and Charles Augustin Coulomb\cite{Coulomb}. They are the independence of friction on the apparent area of contact, the proportionality of the friction to the applied load and the independence of the kinetic friction on the velocity. These laws are concluded from the macroscopic experiments.

In recent years, due to the invention of atomic force microscopy (AFM) by Binnig, Quate and Gerber\cite{AFM}, mechanical properties of materials have been found to have strong size dependence\cite{Bhushan}. Tribological properties could also have size effect on the microscale and nanoscale. The invention of the atomic force microscopy (AFM) by Binnig, Quate and Gerber\cite{AFM} brings us the chances to study mechanical materials in micro- and nanometer scales, in which the studies of friction is defined as micro- and nanotribology.

Several studies\cite{Maboudian}\cite{1998B}\cite{Komvopoulos}\cite{2003B}\cite{B2004} have found that friction and wear properties are the limiting factors to the performance and reliability of the nanoelectromechnical systems (NEMS) devices. And in reality many NEMS devices work at high sliding speeds (m/s) relative to the counterparts, e.g. the I/O-head of a hard disk. However, due to the limitation of the piezo driving stage, the maximum sliding velocity we could get using AFM is less than mm/s. This requires the study of the nanotribology at relative high speeds. Tao et al.\cite{Tao} uses a custom-calibrated ultrahigh velocity stage, developed by Tambe and Bhushan\cite{Tambe}, to produce the high relative speed between the tip and the sample surface. However, the velocity could only get up to 200mm/s.

Due to its piezoelectric effect, quartz crystal resonators have been widely used in research. The amplitude of the oscillation of the quartz crystal was first directly measured by Borovsky \cite{Borovsky} via scanning tunneling microsope. In this article, we report how to directly measure the oscillation amplitude of the quartz crystal with AFM. The oscillation speed is proved to be much higher than mm/s, up to several meters per second. Although such kind of high speed is different from steady or rotary sliding, it is also important to study one fixed position on the sample.Berg et al.\cite{Berg} have recently studied the microtribology with quartz crystals based on ring-down experiments insteady of AFM direct imaging. They found that the friction foce in metal-metal contacts increases approximately linearly with speed when the oscillation is above a minimum critical amplitude.
We combine the quartz crystal oscillating circuits together with the AFM. At the same time we directly image the oscillation of sample surface and make friction measurements. The oscillation speed could get up to 4m/s. In order to demonstrate the utility of this setup, the friction measurements under three different designs are studied.

\section{Experiment}

\begin{figure}[htp]
\centering
\includegraphics[width=8cm]{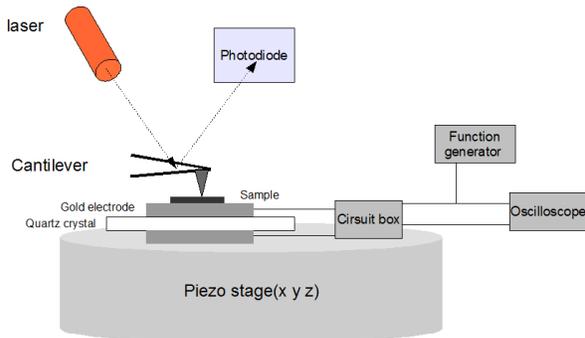}
\caption{\emph{Schematic diagram of the incorporation of the quartz crystal resonator and the commercial AFM: A quartz crystal resonator together with the circuit box is implemented onto the piezo stage. During the oscillation of the quartz crystal, the friction image obtained from AFM shows the tribology under high speed.}}
\label{setup}
\end{figure}

In order to get the relative high speed between the cantilever tip and the sample surface, we use the quartz crystal resonator as the substrate of our samples. By fixing the quartz crystal holder and its oscillating circuits onto the piezo stage [see Fig.\ref{setup}], we modify the commercial AFM setup (DI3100, Nanoscope IIIa controller, Digital Instruments, Santa Barbara, CA) to our purpose. Through the collection of the signal by photodiode, we can get the friction properties at dynamic and static states.

We use the Fischer Pattern on the electrode of the quartz crystal as our sample. The substrate is the quartz crystal resonator (KVG Quartz Crystal Technology GmbH) of 3MHz fundamental mode 'AT-Cut'. It consists a thin blank of a single quartz crystal of diameter 15mm and Gold electrodes of diameter 6mm sputtered onto each side of the disk. The crystal is mounted onto a sample holder with two points of contact, which on the other side connects to the circuit box through thin conductive wires. Together with the circuit box, the quartz crystal holder is tightly fixed onto the AFM piezo stage.

How do we prepare the Fischer Pattern on one of the electrodes? We first evaporate 30 nm SiO$_{x}$ on the gold electrode. This is the prerequisite for the next step. Then we put colloids on the surface. Now the colloids will distribute on the surface. Afterwards, we evaporate 30nm Cr (high adhesion to the surface) onto the surface, which goes into the small areas between the colloids. Finally we remove the colloids by using EMK in Ultrasonic for 1 minutes. Now the Fischer Pattern has been successfully prepared, leaving many triangles on the surface.

\begin{figure}[htp]
\centering
\includegraphics[width=8cm]{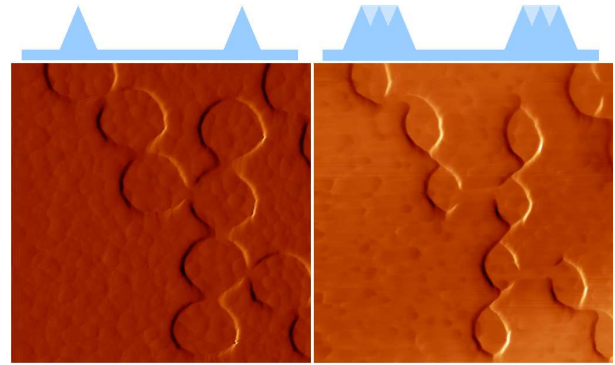}
\caption{\emph{Fischer Pattern on a quartz crystal in static state (left) and in dynamic state (right). The upper drawings are the sketch of two triangels on the base flat layer in both states. The lower images(1.53$\mu$m*1.53$\mu$m) show topographies got by AFM imaging in static state(left) and during oscillation of quartz crystal(right).}}
\label{oscillation}
\end{figure}

We could also use the gold electrode as our sample to measure oscillation amplitude and friction properties, however, because the gold particles sit very close to each other or cover each other, during the oscillation of the quartz crystal, nearest particles will block each other's extension and the cleave in between is smaller or no if averaged over time. Thus AFM tip can not recognize where is the edge of two neighbour particles. Therefore, we could not get the oscillation amplitude by AFM imaging. But with Fischer Pattern on the surface of the electrode, AFM can achieve that. The triangles in the Fischer Pattern are the layer higher than the base layer, for example as shown in Fig.\ref{oscillation}. During oscillation, there is no difficulty for AFM to image the oscillating edge as shown in Fig.\ref{oscillation}.

\begin{figure}[htp]
\centering
\includegraphics[width=8cm]{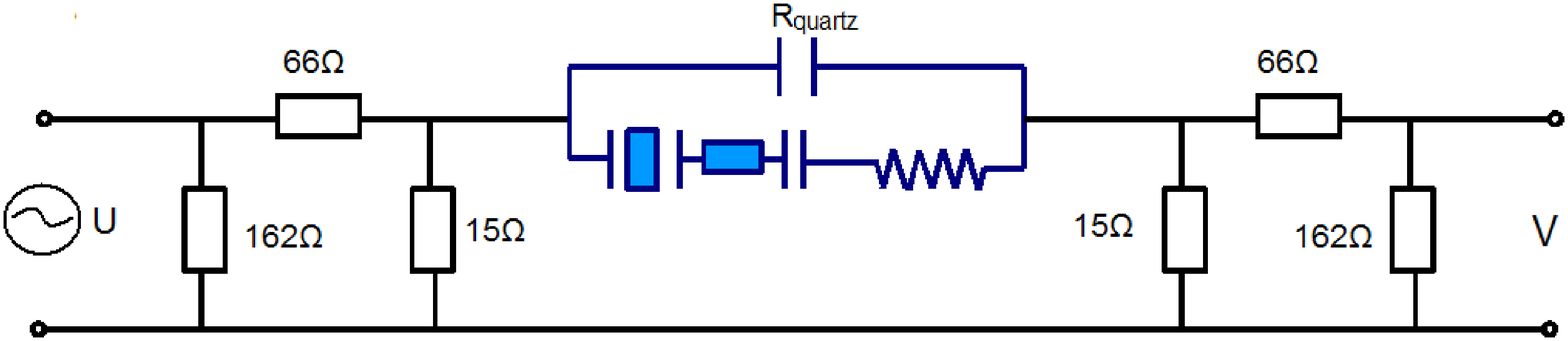}
\caption{\emph{Schematic diagram of the equavalent electrical circuits of the driving system of quartz crystal. }}
\label{circuit}
\end{figure}

Fig.\ref{circuit} shows the equivalent electrical circuit of the quartz crystal and its working electrical box. The blue part is the electrical circuit of the quartz crystal itself. Left and right resistors are symmetrical and used to get rid of the effect of the connecting cable. They are also used to avoid the case of voltage reflection. Function generator is used to oscillate the quartz crystal at certain frequency and certain voltage. We collect the input signal U and output signal V, compare the phase of them. The changing of the driving frequency changes the phase shift between input and output signals. Then the resonance frequency is the one at which input and output signal are of no phase shift and the output signal gets its maximum. The input and output voltages at resonance are read out from oscilloscope as U and V. Then the peak-peak voltage just applied on the quartz crystal can be calculated out through

\begin{equation}
V_{quartz} = 0.1807 \times U-2.6588 \times V
\label{equvolt}
\end{equation}

Before we start the AFM measurement, we turn on the oscillation circuit of the quartz crystal. Attune the function generator to find the resonance frequency of the free oscillation of the quartz. After that, we drive the piezo to bring the tip onto the sample surface. When the tip is approaching the surface, we can see the resonance shift of the oscillation. After the AFM measurement starts, we readjust the driving frequency to get its resonance under applied load. The normal force error image and the friction trace-retrace images can be recorded by AFM software. And after one such measurement, we stop the oscillation. Immediately we make another AFM measurement at the same area with same parameters. The reason why we do this is to keep the same situation and same calibration factor for the setup. Compare the images at dynamic and static state, not only the friction property but also the oscillation amplitude of the quartz crystals can be determined.

\section{Results and discussions}
Before we make friction measurements, we first try to find the areas of interests. As is already established\cite{Martin} that the amplitude of oscillation gets maximum at the center of the electrode and follows Gaussian distribution as the distance from center increases, the AFM tip is located roughly at the center of the quartz crystal. From our experiences, the areas with single triangles far away from each other are easily to be kicked away. The areas with connecting triangles or with local circular lower layers are optimum for the measurements during oscillation. Friction measurements with three different methods have been carried out.

\subsection{}
First make error and friction measurements under different normal loads at static state. Then start to oscillate the quartz crystal. Fix the driving voltage on the function generator and make same AFM measurements as at static mode. So now we can compare the friction property under certain normal loads and with and without high speeds.

\begin{figure}[htp]
\centering
\includegraphics[width=8cm]{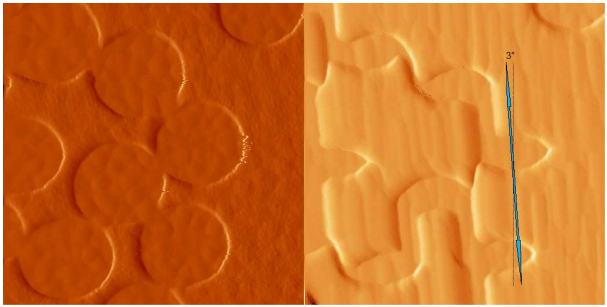}
\caption{\emph{The error images of the Fischer Pattern on the electrode of the quartz crystal at static and dynamic states.  }}
\label{sample1}
\end{figure}

In Fig.\ref{sample1} the left image shows the error signal of the static sample while the right image shows the error signal of the sample oscillated by the quartz crystal at resonance frequency 2997737Hz. All measurements are scanned horizontally by AFM at scanning speed 9.33$\mu$m/s. Compare the images at static and dynamic states, we can conclude the oscillation direction is about 3 degrees to the vertical direction. Using software Image Processing and Analysis in Java (ImageJ\cite{imageJ}) the distances of the half circles in the oscillation direction in both cases can be calculated. Half of the substraction of the distances in both cases gives us the oscillation amplitude.

\begin{equation}
A_{0} = 128 \pm 4.2 nm
\end{equation}
Now we need to get the oscillation speed of the quartz crystal. First let's look at the driving voltage from function generator

\begin{equation}
U_{drive} (t) = U_{0} \times \sin (2 \pi f_{0}t)
\end{equation}

$f_{0}$ is the driving frequency. $V_{0}$ is the amplitude of the driving voltage, in these measurements $V_{0}$ is 10 Vlots.
By equation \ref{equvolt} we can get the peak-peak voltage applied on the quartz crystal is 1.21 V.

Due to the piezoelectric effect of the quartz crystal, the quartz oscillates as

\begin{equation}
a_{quartz} (t) = A_{0} \times \sin (2 \pi ft+\phi)
\label{osequ}
\end{equation}

Here $a_{quartz}$ is the oscillation distance from the static state, $A_{0}$ is the amplitude, f is the oscillation frequency and $\phi$ is the phase shift from the driving source. Before the AFM measurement, the oscillation of quartz crystal has been adjusted to oscillate at its resonance, so

\begin{equation}
f = f_{0} \ and \ \phi = 0
\end{equation}

As velocity is the derivative of equation \ref{osequ}

\begin{equation}
\upsilon (t) = 2 \pi f \times A_{0} \cos (2 \pi ft)
\end{equation}

Therefore the oscillation amplitude is
\begin{eqnarray}
\upsilon_{0} &=& 2 \pi \times 2997737 \times 128 \pm 4.2 \ nm/s \\
                         &=& 2.41 \pm 0.08 m/s
\end{eqnarray}

We use the nondestructive unloaded resonance method of Cleveland \emph{et al.}\cite{Cleveland}\cite{Green} to calibrate the normal force signal. With the resonance frequency of the cantilever of 24.07kHz, the normal force calibration factor is calculated as 111.98nN/V. The lateral force calibration is carried out on the testgrid TGF11\cite{TGF11} using the \emph{Varenber et al.}\cite{Varenberg} wedge calibration method improved from the original \emph{Ogletree et al.}\cite{Ogletree} wedge method. The resultant lateral force calibration factor is about 28.87nN/V.

\begin{figure}[htp]
\centering
\includegraphics[width=8cm]{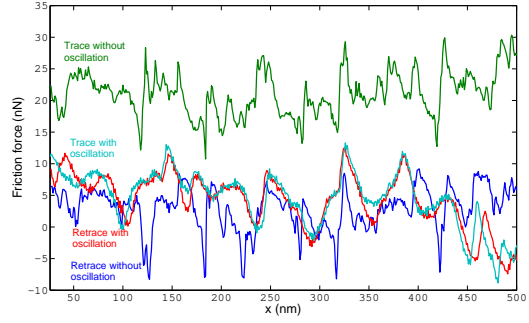}
\caption{\emph{One example of the trace and retrace friction signal measured at static state and during the oscillation of the quartz crysal.}}
\label{linecrosssection}
\end{figure}

Fig.\ref{linecrosssection} shows one line forward scanning and immediately one backward scanning line. As shown in the figure, for the same position on the sample, crosssections with and without quartz oscillation are presented. Clearly, the friction signal which is the subtraction of trace and retrace signals is much larger in the low speed than that in the high speed measurement.

\begin{figure}[htp]
\centering
\includegraphics[width=8cm]{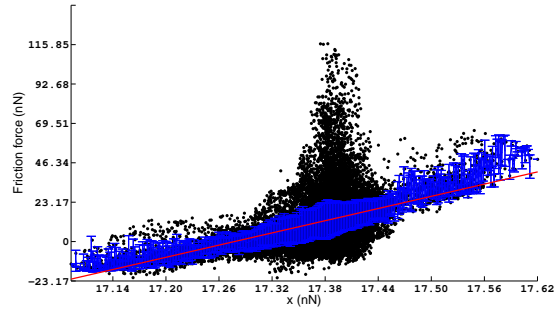}
\caption{\emph{Plot of one AFM friction measurements at applied load 33.59nN. The black dots are the raw data from AFM images, the blue histogram is after binning the raw data in equal small distance of the normal force and the red fitting curve shows the local friction coefficient.}}
\label{binned}
\end{figure}

What is more interesting for us is the friction coefficient for each single measurement. We assume the attractive surface forces such as capillary forces or van der Waals forces are negligible,so the applied load from AFM conbined with error signal is the normal force. To simplify the problem, a linear relation of the friction force on normal force is presumed as shown in the Fig.\ref{binned}. Then we compare the friction coefficients for different cases.

\begin{figure}[htp]
\centering
\includegraphics[width=8cm]{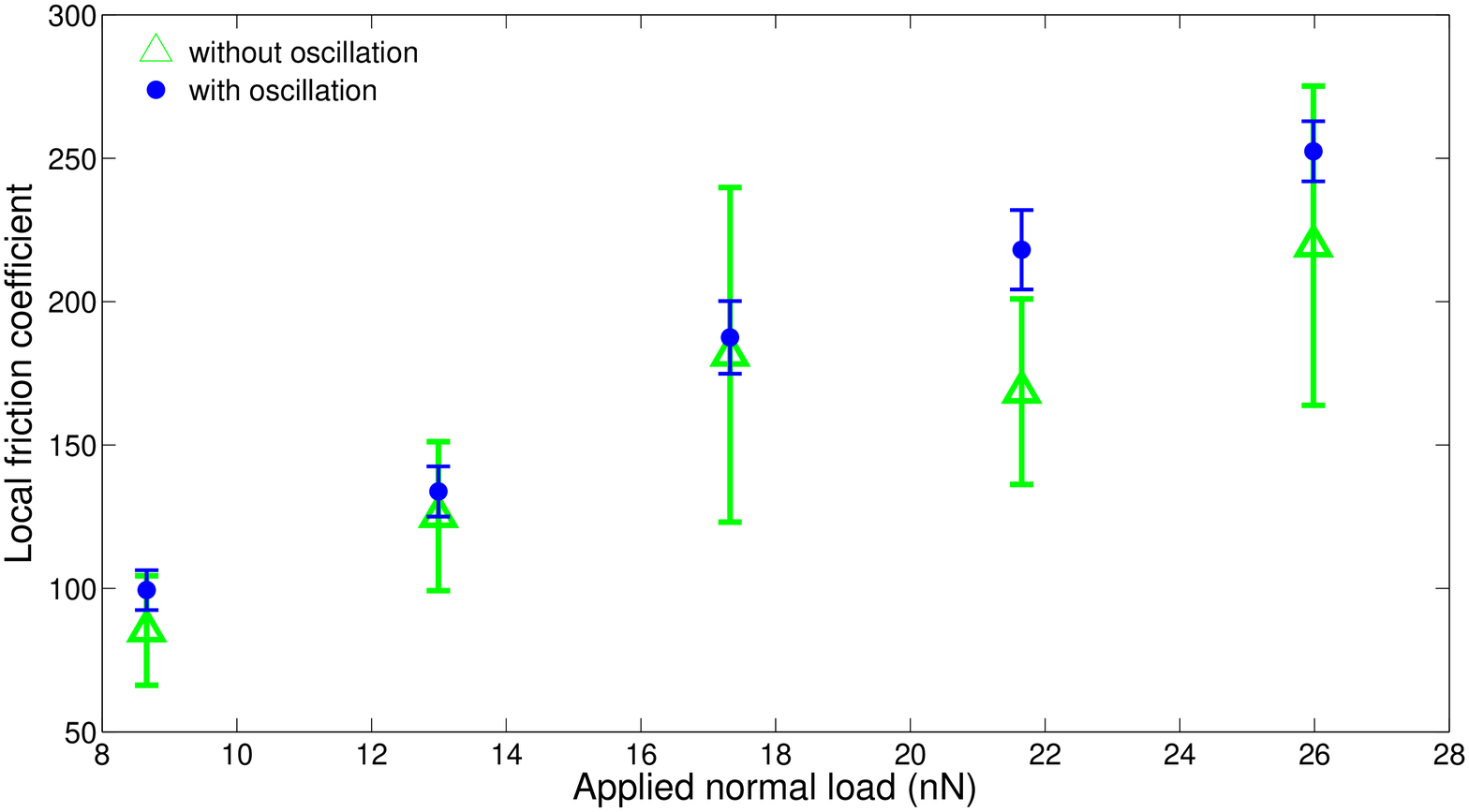}
\caption{\emph{Friction coefficients of the sample surface under different applied normal loads at low scanning speed 9.33$\mu$m/s without oscillation and with oscillation speed 2.41m/s. }}
\label{friction1}
\end{figure}

Fig.\ref{friction1} shows the collective friction coefficients for different AFM setpoints. The blue points are the measured friction coefficients when the quartz crystal are in static state. While the quartz crystal oscillates, the sample particles oscillates together with the crystal, which results in the corresponding green points. Comparison of both cases, the friction coefficients in dynamic state are always higher than that in static state. During the oscillation, the equivalent roughness of the surface is larger than that in static state, because the measured area of the highest layer becomes larger than that without oscillation. The increased roughness could be the reason of the bigger local friction coefficient when the quartz crystal oscillates. According to Amonton's Law, the friction coefficient does not depend on the normal load, which does not fit our case. The rough linear increasing of the friction coefficients on the normal loads denotes that the friction force is dependent on the square of the normal force\ref{equfriction}.

\begin{equation}
\frac{\partial{\emph{f}}}{\partial{N}}=\emph{k}N+b
\end{equation}
\begin{equation}
\emph{f}=\frac{1}{2}\emph{k} N^{2}+\emph{b}N+\emph{a}
\label{equfriction}
\end{equation}

where \emph{f} is the friction force, N is the normal load, \emph{k},\emph{b},\emph{a} are constants.

In Fig.\ref{friction1}, at the value of 33.59nN there are much bigger deviation in the friction force. This is also the set-point of the AFM measurement. The reason could be the confliction of the tip with the oscillating surface. Have a look at the binned data, the standard deviation at this point is much smaller. So the large deviation from the original data could also come from the much more data points at that position.

\subsection{}We make each time under certain normal load one AFM measurement with oscillating quartz crystal and immediately one measurement after stopping the oscillation. Change the normal load and repeat the same measurements. For different normal loads, the oscillation speed is the same.
\begin{figure}[htp]
\centering
\includegraphics[width=8cm]{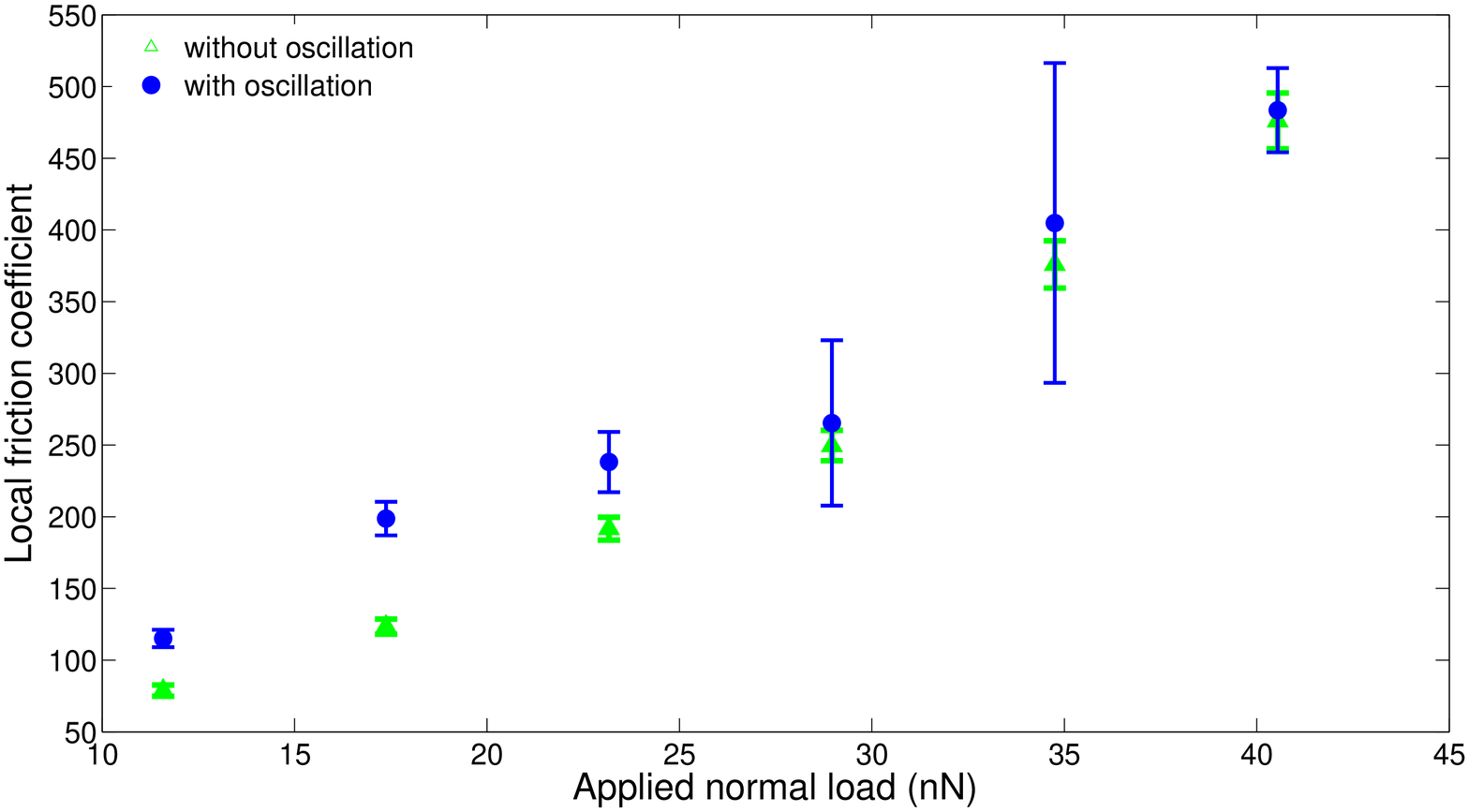}
\caption{\emph{Friction coefficients of the sample surface under different applied normal loads at low scanning speed 20$\mu$m/s without oscillation and with oscillation speed 1.84m/s. }}
\label{friction228012009}
\end{figure}

As shown in Fig.\ref{friction228012009}, there is similar phenomena as in the last method, but the friction coefficients get closer at higher normal loads. The reason could be the instability of the AFM setup. At the beginning of the measurements, the setup is calibrated and the cantilever is new. After some scans, the tip is broadened and there could also be a small movement of the laser spots. However, the tendency of the larger friction coefficients at dynamic state and the increasing of the friction coefficients with increasing normal loads are clear for us.

\subsection{}Still we make one dynamic measurement and one following static measurement. But this time all measurements are under same normal load. We change the driving voltages, which also means that the oscillation speeds are different.
\begin{figure}[htp]
\centering
\includegraphics[width=8cm]{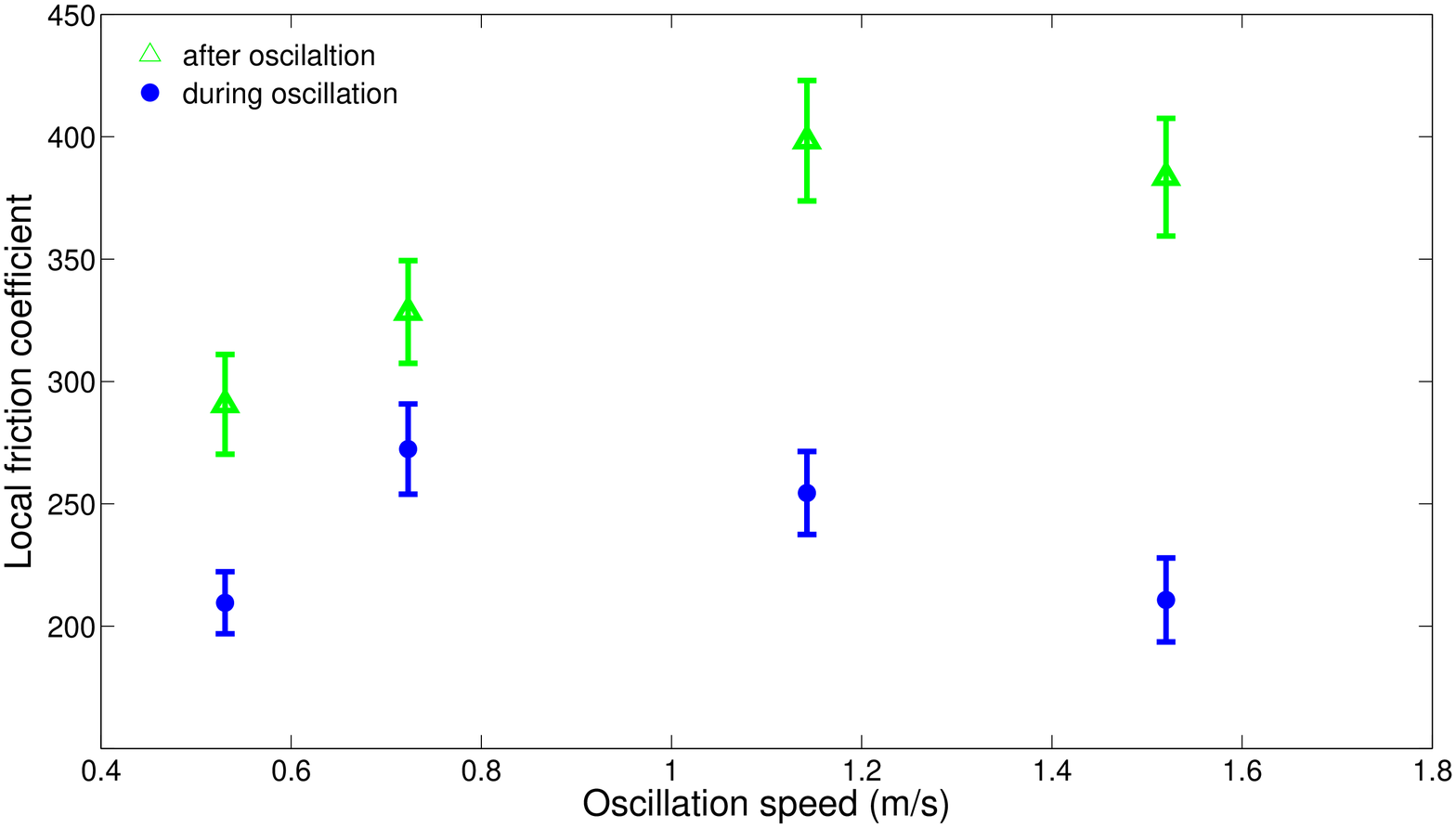}
\caption{\emph{Friction coefficients of the sample surface under 17.3nN normal load  at scanning speed 7.93$\mu$m/s with and without different oscillation speeds.}}
\label{comparspeeds}
\end{figure}

In Fig.\ref{comparspeeds}, the upper blue data present the friction coefficients of the surface under different oscillation speeds. Corresponding friction coefficients after each single oscillation are shown in green. Theoretically, the friction coefficients of same cantilever on same area of the surface with same scanning speed should be the same. However, due to the oscillation, the particles could be rearranged, the tip would be broadened and the AFM setup could be slightly deviated. Therefore, the friction coefficients after each oscillation are changed. Nevertheless, the friction coefficients have the tendency to increase with increasing oscillation speeds.

\section{Conclusions}

We have shown how to study tribology with atomic force microscopy in a range of high speed up to several meter per second. Friction measurements under different designs could be achieved with AFM, no matter in which direction the sample oscillates to the direction of the scanning. Conclude from three different friction measurements, the local friction coefficients in high oscillating speeds are higher than that without oscillation.

\section{Acknowledgement}

This project is financially supported by Landesstiftung Baden-W\"{u}rttemberg. The authors thank Hartmut Schimming for helpful discussions and technical support. The authors would also like to thank M. Gon\c{c}alves and M. Asbach for the help in sample preparation, the Institute of Solid State Physics for using thermal evaporation system.

\bibliographystyle{unsrt}
\bibliography{jabref}

\begin{thebibliography}{10}

\bibitem{Dowson}
D.~Dowson.
\newblock {\em History of Tribology}.
\newblock Longman London, 1979.

\bibitem{Vinci}
Codex~Atlanticus Leonardo~da Vinci and other books.

\bibitem{Amontons}
G.~Amontons.
\newblock {\em Mémoires de l'académie royale}, A:257, 1706.

\bibitem{Coulomb}
C.~A. Coulomb.
\newblock Théorie des machines simples, en ayant égard au frottement de leurs
  parties, et à la roideur des cordages.
\newblock {\em Mém. Math. Phys.}, page 161, 1785.

\bibitem{AFM}
C.F.~Quate G.~Binnig and Ch. Gerber.
\newblock Atomic force microscope.
\newblock {\em Phys. Rev. Lett.}, 56 (9):930--933, 1986.

\bibitem{Bhushan}
B.~Bhushan.
\newblock {\em Springer Handbook of Nanotechnology}.
\newblock 2nd ed. (Springer, Heidelberg, Germany), 2006.

\bibitem{Maboudian}
R.~Maboudian and R.T. Howe.
\newblock Critical review: Adhesion in surface micro-mechanical structures.
\newblock {\em J Vac Sci Technol}, B:15:1--20, 1997.

\bibitem{1998B}
Bhushan B.
\newblock {\em Tribology issues and opportunities in MEMS}.
\newblock Kluwer, Dordrecht, 1998.

\bibitem{Komvopoulos}
K.~Komvopoulos.
\newblock Adhesion and friction forces in microelectromechanical systems:
  Mechanisms, measurement, surface modification techniques, and adhesion
  theory.
\newblock {\em J Adhes Sci Technol}, 17 (4):477--518, 2003.

\bibitem{2003B}
Bhushan B.
\newblock {\em J Vac Sci Technol}, B 21:2262--2296.

\bibitem{B2004}
Bhushan B.
\newblock {\em Springer handbook of nanotechnology}.
\newblock Springer, Heidelberg, Germany, 2004.

\bibitem{Tao}
Zhenhua Tao and Bharat Bhushan.
\newblock New technique for studying nanoscale friction at sliding velocities
  up to 200mm/s using atomic froce microscope.
\newblock {\em Rev. Sci. Instrum.}, 77(10):103705, 2006.

\bibitem{Tambe}
Nikhil~S Tambe and Bharat Bhushan.
\newblock A new atomic force microscopy based technique for studying nanoscale
  friction at high sliding velocities.
\newblock {\em J. Phys. D: Appl. Phys.}, 38:764~773, 2005.

\bibitem{Borovsky}
B.~L.~Mason B.~Borovsky and J.~Krim.
\newblock Scanning tunneling microscope measurements of the amplitude of
  vibration of a quartz crystal oscillator.
\newblock {\em J. Appl. Phys.}, 88(7):4017~4021, 2000.

\bibitem{Berg}
S.~Berg and D.~Johannsmann.
\newblock High speed microtribology with quartz crystal resonators.
\newblock {\em Phys. Rev. Lett.}, 91 (14):145505--1~145505--4, 2003.

\bibitem{Martin}
B.~A. Martin and H.~E. Hager.
\newblock Velocity profile on quartz crystals oscillating in liquids.
\newblock {\em J. Appl. Phys.}, 65:7, 1989.

\bibitem{imageJ}
http://rsbweb.nih.gov/ij/.

\bibitem{Cleveland}
J.~P. Cleveland and P.~K.~Hansma S.~Manne, D.~Bocek.
\newblock A nondestructive method for determining the spring constant of
  cantilevers for scanning force microscopy.
\newblock {\em Rev. Sci. Instrum.}, 64(2):403--405, 1993.

\bibitem{Green}
Jason P.~Cleveland Christopher P.~Green, Hadi~Lioe and John E.~Sader
  Roger~Proksch, Paul~Mulvaney.
\newblock Normal and torsional spring constants of atomic force microscope
  cantilevers.
\newblock {\em Review of scientific instruments}, 75(6):1988--1996, 2004.

\bibitem{TGF11}
http://www.spmtips.com/tgf.

\bibitem{Varenberg}
I.~Etsion M.~Varenberg and G.~Halperin.
\newblock An improved wedge calibration method for lateral force in atomic
  force microscopy.
\newblock {\em Review of scientific instruments}, 74(7):3362--3367, 2003.

\bibitem{Ogletree}
R.~W.~Carpick D.~F.~Ogletree and M.~Salmeron.
\newblock Calibration of frictional forces in atomic froce microscopy.
\newblock {\em Rev. Sci. Instrum.}, 67(9):3298--3306, 1996.

\end{thebibliography}

\end{document}